\documentclass[twocolumn,showpacs,preprintnumbers,amsmath,amssymb]{revtex4}

\usepackage{graphicx}% Include figure files
\usepackage{dcolumn}% Align table columns on decimal point
\usepackage{bm}% bold math
\usepackage{bezier}

\usepackage{amsmath}
\usepackage{enumerate}
\usepackage{hyperref}
\usepackage{subfigure}
\usepackage{epstopdf}
\usepackage{multi row}

\begin{document}
\title{Relaxation and Thermalization of Isolated Many-Body Quantum Systems}

\author{E. J. Torres-Herrera}
\author{Davida Kollmar}
\author{Lea F. Santos}
\affiliation{Department of Physics, Yeshiva University, New York, New York 10016, USA}

\date{\today}

\begin{abstract}
We provide an overview of our numerical and analytical studies of isolated interacting quantum systems that are quenched out of equilibrium instantaneously. We describe the relaxation process to a new equilibrium and obtain lower bounds for the relaxation time of full random matrices and realistic systems with two-body interactions. We show that the size of the time fluctuations after relaxation decays exponentially with system size for systems without too many degeneracies. We also discuss the conditions for thermalization and demonstrate that it can happen after local and global quenches in space. The analyses are developed for systems, initial states, and few-body observables accessible to experiments with optical lattices.
\end{abstract}


\maketitle

%%%%%%%%%%%%%%%% INTRODUCTION %%%%%%%%%%%%%%%%%%%%%%%%%
\section{Introduction}

The unitary evolution of many-body quantum systems is an active area of research with studies being carried out theoretically and also experimentally. In solid state nuclear magnetic resonance~\cite{Cappellaro2007,Ramanathan2011,Kaur2013}, and especially in experiments with optical lattices~\cite{Chen2011,Aidelsburger2011,Trotzky2012,Trotzky2008,Simon2011,Greif2013,Fukuhara2013,HildARXIV}, the effects of the environment take a long time to become relevant, which allows for treating the systems as quasi-isolated. In this context, much attention has been given to the subject of quench dynamics, where 
the evolution of an isolated quantum system initiates with an instantaneous perturbation that takes a certain initial Hamiltonian $\widehat{H}_I$  into a new final Hamiltonian $\widehat{H}_F$.

Studies of quenched many-body quantum systems deal with fundamental problems of physics, such as advancing our understanding of nonequilibrium quantum physics and deriving thermodynamics from quantum mechanics. Among the several specific questions that have been raised, we address three that we have been particularly interested in~\cite{Zangara2013,Torres2013,Torres2014PRA,Torres2014NJP,Torres2014PRAb,Torres2014PRE}. 

(i) How fast can isolated many-body quantum systems evolve? We try to answer this question by studying the quantum fidelity decay and the evolution of few-body observables. 

The quantum fidelity corresponds to the overlap between two quantum states, in our case the initial state and its evolved counterpart. For this choice, the fidelity is obtained by Fourier transforming the weighted energy distribution of the initial state. This distribution is often called local density of states (LDOS) or strength function. The fidelity considered here is also related to the Loschmidt echo and the characteristic function of the probability distribution of work~\cite{Silva2008,Silva2009,Gambassi2011}, which is essential in studies about work statistics in quantum thermodynamics.

The minimum fidelity decay time was derived from the time-energy uncertainty relation~\cite{Mandelstam1945,Bhattacharyya1983,Pfeifer1993,Uffink1993,GiovannettiPRA2003,Giovannetti2003,Giovannetti2004}. This limit can be reached in particular scenarios, such as the one we describe in Ref.~\cite{Torres2014PRAb}, where the LDOS of the initial state is bimodal. In the general picture where the LDOS is single-peaked, we show that the fastest possible decay occurs for evolutions under full random matrices~\cite{Torres2014PRA,Torres2014NJP,Torres2014PRAb}. 

Full random matrices provide a way to acquire statistical information about the spectrum of complex systems, but they are unphysical, because they imply the simultaneous interactions of many particles. In realistic systems with two-body interactions, the fastest fidelity decay is Gaussian and this behavior may hold until saturation~\cite{Torres2014PRA,Torres2014NJP,Torres2014PRAb}. The Gaussian decay happens when the shape of the LDOS is also Gaussian and it fills the energy shell. The latter gives the maximum possible spreading of initial states projected onto final two-body-interaction Hamiltonians.  

The evolution of few-body observables depends not only on the interplay between the initial state and final Hamiltonian, as the fidelity, but also on the details about the observable, which makes the study more difficult. However, if the observable commutes with the initial Hamiltonian, a simple picture emerges at short times. In this case, the dynamics is quadratic in time.

(ii) How large are the time fluctuations of observables around the steady state after relaxation? Equilibration in isolated quantum systems can happen in a probabilistic sense. It requires that the time fluctuations be  small, implying proximity to the stationary state for the vast majority of time, and that the fluctuations decrease with system size, vanishing in the thermodynamic limit.  We show that the temporal fluctuations of the fidelity decrease exponentially with system size. This is in agreement with our analyses of the fluctuations of few-body observables and holds for chaotic and also integrable systems, provided the spectrum be not highly degenerate~\cite{Zangara2013}.

(iii) After relaxation, what are the conditions for the new reached equilibrium to be thermal? An observable is said to have thermalized when its infinite time average coincides with its thermal average. For a final two-body-interaction Hamiltonian, integrable or chaotic, thermalization occurs when the LDOS fills the energy shell ergodically, for example when the initial state is an eigenstate from a full random matrix.
But this is not a very interesting case, since such initial states have infinite temperature and are already thermal before the quench~\cite{Torres2013}. For the realistic scenario of initial states with a finite temperature, and also with a narrow spreading in energy, thermalization should be viable when the eigenstates of the final Hamiltonian are chaotic and the energy of the initial state is away from the edges of the spectrum of $\widehat{H}_F$ (the borders are problematic, because there the states are not strongly mixed for systems with few-body interactions) \cite{Jensen1985,Deutsch1991,Srednicki1994,Horoi1995,ZelevinskyRep1996,Flambaum1996b,Flambaum1997,Borgonovi1998,Borgonovi2000,I01,Rigol2008,rigol09STATa,rigol09STATb,Santos2010PRE,RigolSantos2010,Santos2010PREb,Santos2011PRL,Rigol2012,He2012,He2013,Torres2013,Torres2014PRE}. We provide illustrations supporting this idea. 

We show that the infinite time average of few-body observables and their thermodynamic average approach each other as the energy of the initial state moves closer to the center of the spectrum of $\widehat{H}_F$, as the perturbation gets stronger and the eigenstates more chaotic, and as the system size increases. This holds for global and also local quenches in space~\cite{Torres2014PRE}.

This paper is organized as follows. Section~\ref{Sec:dos} describes the models.
Section \ref{Sec:ldos} analyzes the relationship between fidelity decay and the LDOS. 
Section \ref{Sec:Fluctuations} gives results for the temporal fluctuations after relaxation.
The viability of thermalization is discussed and illustrated in Sec.~\ref{Sec:Thermalization}. Concluding remarks are presented in Sec.~\ref{Sec:Summary}.

%%%%%%%%%%%%%%%% MODEL %%%%%%%%%%%%%%%%%%%%%%%%%
\section{Model and Quench}
\label{Sec:dos}

Full random matrices are often employed to obtain statistical information about the spectrum of many-body quantum systems. The matrices are filled with random numbers and satisfy the symmetries of the system they aim at describing. In time-reversal invariant systems with rotational symmetry, the matrices are real and symmetric, as the ones considered in this paper. They belong to the so-called Gaussian orthogonal ensemble (GOE). The level spacing distributions of full random matrices show level repulsion~\cite{Wigner1957,Wigner1967,HaakeBook,Guhr1998}. However, despite providing statistical properties in agreement with the spectra of complex systems, full random matrices are unrealistic, because they imply the simultaneous interactions of many particles. Realistic systems have few-body interactions. These systems also show level repulsion when they are in the chaotic domain, but other properties differ from full random matrices. For example, the density of states of full random matrices has a semicircular shape and all of their eigenstates are pseudo-random vectors, whereas in realistic systems with two-body interactions the density of states is Gaussian and the eigenstates can only reach a high level of delocalization in the middle of the spectrum, being more localized at the edges~\cite{ZelevinskyRep1996,Brody1981,Izrailev1990,Gubin2012}. 

\subsection{Spin-1/2 Model}
\label{Sec:model}
The realistic systems that we investigate correspond to spin-1/2 models with two-body interactions and no randomness. These models describe real magnetic compounds~\cite{Sologubenko2000PRB,Hess2007,Hlubek2010}, crystals of fluorapatite~\cite{Cappellaro2007,Ramanathan2011,Kaur2013}, and have also been simulated with optical lattices~\cite{Simon2011,Greif2013,Fukuhara2013,HildARXIV}. We focus on one-dimensional lattices with open boundaries and an even number $L$ of sites. They can be mapped onto systems of spinless fermions~\cite{Jordan1928} or hardcore bosons~\cite{Holstein1940}. The Hamiltonian is given by
\begin{eqnarray}
&& \widehat{H} = \varepsilon J \widehat{S}_{1}^z  + d J \widehat{S}_{\lfloor L/2 \rfloor }^z + \widehat{H}_{NN} + \lambda \; \widehat{H}_{NNN} \;,
\label{ham} \\
&& \widehat{H}_{NN} = \sum_{j=1}^{L-1} J \left(\widehat{S}_j^x \widehat{S}_{j+1}^x + \widehat{S}_j^y \widehat{S}_{j+1}^y +\Delta \widehat{S}_j^z \widehat{S}_{j+1}^z \right) \;,
\nonumber \\
&& \widehat{H}_{NNN} = \sum_{j=1}^{L-2} J \left(\widehat{S}_j^x \widehat{S}_{j+2}^x + \widehat{S}_j^y \widehat{S}_{j+2}^y +\Delta \widehat{S}_j^z \widehat{S}_{j+2}^z \right) \;.
\nonumber 
\end{eqnarray}
Above, $\hbar =1$, $\widehat{S}^{x,y,z}_i $ are spin operators acting on site $i$, $\widehat{S}_i^x \widehat{S}_{i+1}^x + \widehat{S}_i^y \widehat{S}_{i+1}^y$ $(\widehat{S}_i^x \widehat{S}_{i+2}^x + \widehat{S}_i^y \widehat{S}_{i+2}^y)$ is the flip-flop term, and $\widehat{S}_i^z \widehat{S}_{i+1}^z (\widehat{S}_i^z \widehat{S}_{i+2}^z)$ is the Ising interaction between NN (NNN) spins. All the parameters, $J,\Delta,\lambda,\varepsilon$ and $d$, are assumed positive. $J$ is the exchange coupling constant, $\Delta$ is the anisotropy parameter (we are interested in the gapless regime, where $\Delta<1$), and $\lambda$ refers to the ratio between NNN and NN couplings.  The total spin in the $z$-direction, $\widehat{{\cal{S}}}^z=\sum_i\widehat{S}_i^z$, is conserved, so the total Hamiltonian is divided into  $\widehat{{\cal{S}}}^z$ subspaces, each with a certain dimension ${\cal D}$. 

The impurities $\varepsilon J$ on the first site of the chain and $d J$ on site $\lfloor L/2 \rfloor$ are generated by applying two local static magnetic fields in the $z$-direction. The purpose of the small defect (impurity) on the first site is to break trivial symmetries, such as parity, conservation of total spin, and spin reversal, without breaking the integrability of the system~\cite{Alcaraz1987}. If $\varepsilon=d=0$, we refer to the system as clean.

Depending on the values of the parameters $\Delta, d$, and $\lambda$, the chain may be integrable or chaotic:

$\bullet $ {\em Integrable XX model}: $d, \Delta, \lambda =0$. This Hamiltonian can be mapped onto a system of noninteracting spinless fermions, being trivially solvable~\cite{Jordan1928}. 

$\bullet $ {\em Integrable XXZ model}: $\Delta\neq 0$ and $d, \lambda=0$ . This model is solved by means of the Bethe ansatz~\cite{Bethe1931}.  

$\bullet $ {\em Chaotic impurity model}: $\Delta, d \neq 0$ and $\lambda=0$. The addition to the XXZ Hamiltonian of a single impurity close to the middle of the chain can bring the system into the chaotic domain~\cite{Santos2004,Barisic2009,Santos2011,Torres2014PRE} provided $d \lesssim 1$. If the defect becomes too large it splits the system in two independent and integrable chains. The onset of chaos is caused by the interplay between the Ising interaction and the impurity. In contrast, the addition of $d$ to the XX model does not affect its integrability. 

$\bullet $ {\em Chaotic NNN model}: $\Delta, \lambda \neq 0 $ and $d=0$. The addition of couplings between second neighbors breaks integrability~\cite{Kudo2004,Kudo2005,Gubin2012,Torres2013}.   

Note that the values of $d$ and $\lambda$ leading to chaos depends on the system size. The larger the system, the smaller the parameter needs to be~\cite{Santos2010PRE,Torres2014PRE}.

%%%%%%%%%%%%%%%% QUENCH and NUMERICAL METHOD %%%%%%%%%%%%%%%%%%%%%%%%%
\subsection{Quench Dynamics}
\label{Sec:quench}

In the scenario of quench dynamics, the initial state  $|\Psi(0)\rangle = |\text{ini}\rangle$ is an eigenstate of the initial Hamiltonian $\widehat{H}_\text{I}$. The dynamics starts with the sudden change of some parameter(s) of this Hamiltonian in a time interval much shorter than any characteristic time scale of the model. It brings the system to the final Hamiltonian $\widehat{H}_\text{F}$ with eigenvalues $E_\alpha$ and eigenstates $|\psi_\alpha\rangle$.  

%It is appropriate to study the properties of the final Hamiltonian matrix and of its eigenstates in the basis corresponding to the eigenstates of  $\widehat{H}_\text{I}$. The off-diagonal elements of $\widehat{H}_\text{F}$ provide the number of states directly coupled with the initial state.

In this paper, $J$ sets the energy scale and $\varepsilon$ is fixed. The initial parameters $(\Delta_I, \lambda_I, d_I)$ are the ones that may be quenched into $(\Delta_F, \lambda_F, d_F)$. %We only write the index $I$ or $F$ for parameters that get changed.

The unitary time evolution of the initial state is given by
\begin{equation}
|\Psi(t)\rangle=e^{-i\widehat{H}_\text{F}t}| \text{ini} \rangle=\sum_{\alpha} C_{\alpha}^{\text{ini}}  e^{-iE_\alpha t}|\psi_\alpha\rangle ,
\label{eq:instate}
\end{equation} 
where the coefficients $C_{\alpha}^{\text{ini}} = \langle \psi_{\alpha} | \text{ini} \rangle $  are the overlaps of the initial state with the eigenstates of $\widehat{H}_\text{F}$. 

The evolution is computed numerically with full exact diagonalization for matrices of dimension ${\cal D} < 20\,000$ and with EXPOKIT~\cite{Expokit,Sidje1998} for larger  sizes. 
EXPOKIT is a software package based on Krylov subspace projection methods.
Instead of diagonalizing the complete system Hamiltonian, the package
 computes directly the action of the matrix exponential $e^{-i\widehat{H}_\text{F}t}$ on a vector
of interest.

%%%%%%%%%%%%%%%% LDOS %%%%%%%%%%%%%%%%%%%%%%%%%

\section{Fidelity Decay and Observables Evolution}
\label{Sec:ldos}

The fidelity, also known as survival probability, non-decay probability, or return probability, gives the probability of finding the system still in the initial state after time $t$. It is given by the overlap,
\begin{eqnarray}
F(t) &\equiv&  |\langle \Psi(0)| \Psi(t) \rangle |^2 =  \left| \langle \text{ini} | e^{-i \widehat{H}_F t} |\text{ini} \rangle \right|^2  \\
&=&  \left|\sum_{\alpha=1}^{{\cal D}} |C_{\alpha}^{\text{ini}} |^2 e^{-i E_{\alpha} t}  \right|^2 
\approx  \left|  \int_{-\infty}^{\infty} P^{\text{ini}}(E) e^{-i E t} dE \right|^2 \nonumber.
\label{eq:fidelity}
\end{eqnarray}
In the last term above, the sum was substituted by an integral, which is a good approximation when ${\cal D}$ is large and the density of states is dense. $P^{\text{ini}}(E)$ is the envelope of the LDOS. The latter is obtained numerically by dividing the spectrum of $\widehat{H}_F$ in small windows of energy and computing the sum $\sum_{\alpha} |C_{\alpha}^{\text{ini}} |^2$ inside each window.
One can see that the fidelity is simply the Fourier transform of $P^{\text{ini}}(E)$. Finding an expression for $F(t)$ therefore reduces to identifying the shape of the LDOS.

\subsection{Fastest Fidelity Decay: Semicircular LDOS}
\label{sec:SC}

For an initial state projected onto a full random matrix, $P^{\text{ini}}(E)$ has a semicircular form~\cite{Torres2014PRA,Torres2014NJP,Torres2014PRAb}, 
reflecting the density of states of those matrices. The fidelity for this distribution involves the Bessel function of the first kind, ${\cal J}_1$. We have
\begin{eqnarray}
&&P^{\text{ini}}_{SC}(E)= \frac{2}{\pi {\cal E}} \sqrt{1 - \left(\frac{E}{{\cal E}}\right)^2} ,\nonumber \\
&&F_{\text{SC}}(t) \!=\! \left| \int_{-{\cal E}}^{{\cal E}} P^{\text{ini}}_{SC}(E) e^{-i E t} dE
\right|^2 \!=\! \frac{[{\cal J}_1( 2 \sigma_{\text{ini}} t)]^2}{\sigma_{\text{ini}}^2 t^2},
\label{Ffrm}
\end{eqnarray}
where  $2{\cal E}$ is the length of the spectrum and
%\[
$\sigma_{\text{ini}} = \sqrt{\sum_{\alpha} |C_{\alpha}^{\text{ini}} |^2 E_{\alpha}^2} ={\cal E}/2$
%\]
is the uncertainty in energy of the initial state. An illustration for the semicircular LDOS and its corresponding fidelity decay is 
provided in Fig.~\ref{fig:LDOS} (a) and (b), respectively.

Equation~(\ref{Ffrm}) gives the fastest fidelity decay for a single-peaked LDOS. It is not realistic, since full random matrices are non-physical, but it serves to provide a lower bound. Notice, however, that $F_{\text{SC}}(t)$ is still slower than the ultimate lower bound, $F(t)\geq \cos^2(\sigma_{\text{ini}} t)$, derived from the time-energy uncertainty relation \cite{Mandelstam1945,Bhattacharyya1983,Pfeifer1993,Uffink1993,GiovannettiPRA2003,Giovannetti2003,Giovannetti2004}. This ultimate bound can be approached when the LDOS involves at least two peaks and they are well separated in energy, as we showed in \cite{Torres2014PRAb}. 

After relaxation and in the absence of too many degeneracies, the fidelity, 
\begin{equation}
F(t) = \sum_{\alpha} |C_{\alpha}^{\text{ini}} |^4 + \sum_{\alpha \neq \beta} |C_{\alpha}^{\text{ini}} |^2 |C_{\beta}^{\text{ini}} |^2 e^{i (E_{\alpha} - E_{\beta}) t} ,
\label{eq:Fevolved}
\end{equation}
saturates to its infinite time average,
\begin{equation}
\overline{F}= \lim_{t\rightarrow \infty} \frac{1}{t}
\int^{t}_0 d\tau\,  F(\tau) = \sum_{\alpha} |C_{\alpha}^{\text{ini}} |^4=\frac{1}{\text{IPR}_{\text{ini}}},
\label{eq:Fsaturated}
\end{equation} 
where  the inverse participation ratio, $\text{IPR}_{\text{ini}} $, measures the level of delocalization of the initial state in the energy eigenbasis. A large $\text{IPR}_{\text{ini}} $ indicates a delocalized state. The largest values occur for eigenstates of full random matrices or for a state projected onto such eigenstates. When $\widehat{H}_F$ comes from a GOE, $\text{IPR}_{\text{ini}} \sim {\cal D}/3$ \cite{Izrailev1990,ZelevinskyRep1996}. 

With the analytical expressions in Eqs.~(\ref{Ffrm}) and (\ref{eq:Fsaturated}), we can compute the time    $t_R$ that it takes for the fidelity to first reach the saturation point,
\begin{equation}
\frac{ [{\cal J}_1( 2 \sigma_{\text{ini}} t_R)]^2}{\sigma_{\text{ini}}^2 t_R^2} = \frac{3}{{\cal D}} .
\label{eq:tRfrm}
\end{equation}
The value of $t_R$ obtained from Eq.~(\ref{eq:tRfrm}) gives the lower bound for the relaxation time of many-body quantum systems with a single-peaked LDOS.

After reaching $\overline{F}$, the fidelity shows small oscillations. The envelope of these oscillations is well fitted with $\dfrac{3}{10  (\sigma_{\text{ini}} t)^3}$.

\subsection{Exponential Fidelity Decay: Lorentzian LDOS}

In realistic systems with two-body interactions, the density of states is Gaussian instead of semicircular. In this case, the shape of the LDOS depends of the strength of the perturbation and on the energy, 
\begin{equation}
E_{\text{ini}} = \langle \text{ini} |\widehat{H}_F | \text{ini} \rangle = \sum_{\alpha} |C_{\alpha}^{\text{ini}}|^2 E_{\alpha} ,
\label{Eini}
\end{equation}
of the initial state. For $E_{\text{ini}}$ close to the middle of the spectrum, as the strength of an instantaneous global perturbation applied on $\widehat{H}_\text{I}$ increases from zero, the LDOS broadens from a delta function to a Lorentzian form, and eventually reaches a Gaussian shape. Closer to the edges of the spectrum, the distributions are skewed~\cite{Torres2013,Torres2014PRAb}.

A Lorentzian distribution leads to the exponential fidelity decay~\cite{expRef,Cerruti2002,FlambaumARXIV,Flambaum2000A,Flambaum2001a,Flambaum2001b,Weinstein2003,Emerson2002},
\begin{eqnarray}
&&P^{\text{ini}}_{L}(E) = \frac{1}{2\pi} \frac{\Gamma_{\text{ini}} }{(E_{\text{ini}} - E)^2 +
 \Gamma_{\text{ini}}^2 /4} ,\nonumber \\
&&F_{\text{L}}(t) = \left| \int_{-\infty}^{\infty} P^{\text{ini}}_{BW}(E) e^{-i E t} dE
\right|^2
= e^{- \Gamma_{\text{ini}}  t},
\label{Fbw}
\end{eqnarray}
where $\Gamma_{\text{ini}} $ is the width of the distribution. This is the scenario of the Fermi golden rule.

\subsection{Gaussian Fidelity Decay: Gaussian LDOS}

In the limit of strong perturbation, when the LDOS becomes a Gaussian of width $\sigma_{\text{ini}} $, the fidelity decay is Gaussian~\cite{Flambaum2001a,Flambaum2001b,Izrailev2006,Torres2014PRA,Torres2014NJP,Torres2014PRAb},
\begin{eqnarray}
&& P^{\text{ini}}_{G}(E) = \frac{1}{ \sqrt{ 2 \pi \sigma^2_{\text{ini}} }  } \exp \left[ -\frac{(E-E_{\text{ini}})^2}{ 2 \sigma^2_{\text{ini}} }   \right] \nonumber \\
&&F_{\text{G}}(t) = \left| \int_{-\infty}^{\infty} P^{\text{ini}}_{G}(E) e^{-i E t} dE
\right|^2 = e^{- \sigma_{\text{ini}}^2  t^2}.
\label{eq:Fgauss}
\end{eqnarray}
The Gaussian envelope $P^{\text{ini}}_{G}(E) $ of the LDOS is known as the energy shell. It gives the maximum possible spreading of $|\text{ini} \rangle $ in the eigenstates of the final Hamiltonian. Not all initial states can fill it. The ergodic filling of the energy shell is used as a definition of chaotic states. 

{\bf \em The Gaussian fidelity decay can hold until saturation~\cite{Torres2014PRA,Torres2014NJP,Torres2014PRAb}. The minimum $t_R$ for realistic systems with two-body interactions and a single-peaked LDOS is therefore,}
\begin{equation}
\exp \left( - \sigma_{\text{ini}}^2  t_R^2 \right) =\text{IPR}_{\text{ini}}^{-1} \Rightarrow t_R=\frac{\sqrt{\ln(\text{IPR}_{\text{ini}}) }}{\sigma_{\text{ini}} } .
\end{equation}
The minimum time for $F(t)$ to reach the saturation point is determined by the level of delocalization of the initial state and the width of the energy shell.

\begin{figure}[htb]
\centering
\includegraphics*[width=3.4in]{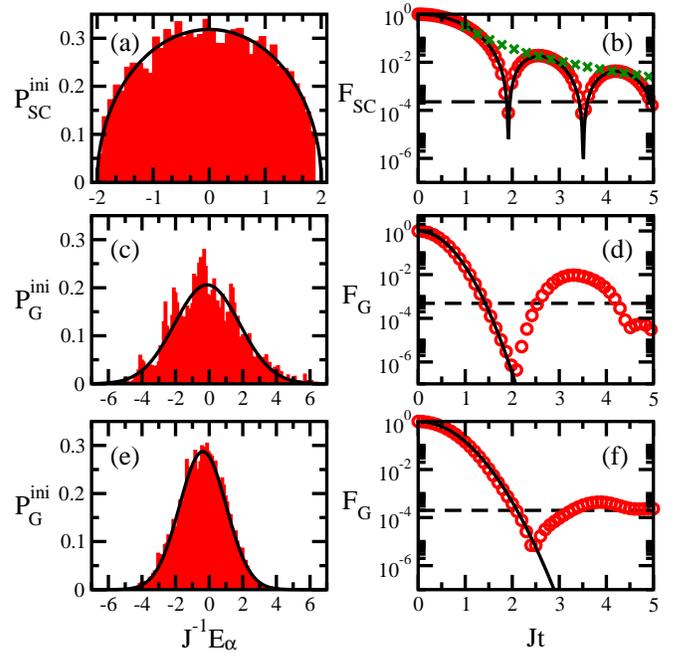}
\caption{(Color online) 
Local density of states (left) and fidelity decay (right). Top: Initial state from a GOE full random matrix projected onto another GOE full random matrix, ${\cal D} = 12\,870$.  The random numbers are normalized so that the length of the spectrum is $4J$ and $\sigma_{\text{ini}}= J$.
Middle: N\'eel state with $E_{\text{ini}}=-J/8$, $\sigma_{\text{ini}}=\sqrt{15}J/2$; $\varepsilon=0,d_F=0,\Delta_F=0.5,\lambda_F=1$, ${\cal \widehat{S}}^z=0$,  $L=16$.
Bottom: Initial state from the XX model with  $E_{\text{ini}}=-0.378J$, $\sigma_{\text{ini}}=1.389J$; $\varepsilon=0.1,d_F=0,\Delta_F=0.48,\lambda_F=1$, ${\cal \widehat{S}}^z=-3$,  $L=18$.  Solid lines give the analytical expressions, shaded area (left) and circles (right) are numerical results. The saturation value of the fidelity is indicated with the dashed horizontal line. In panel (b): crosses indicate the envelope of the oscillations, fitted with $3/[10 (\sigma_{\text{ini}} t)^3]$.}
\label{fig:LDOS}
\end{figure}

Illustrations for $P^{\text{ini}}_{G}(E) $ and $F_{\text{G}}(t)$ are given in Figs.~\ref{fig:LDOS} (c) and (d), respectively. We choose for $|{\text{ini}}\rangle$, the N\'eel state, $|\downarrow \uparrow \downarrow \uparrow  \ldots  \downarrow \uparrow \downarrow  \uparrow \rangle$, whose dynamics can be studied experimentally with optical lattices. We let it evolve according to a clean chaotic final Hamiltonian with $\varepsilon=0, d_F=0,\Delta_F=0.5,\lambda_F=1$. Notice that the Gaussian decay of the fidelity persists until saturation.

Initial states such as the N\'eel state, where each excitation is confined to a single site, constitute the site basis, also known as computational basis or natural basis. They are the eigenstates of the Ising interaction of the Hamiltonian (\ref{ham}), which corresponds then to $\widehat{H}_\text{I}$. 

For site-basis vectors, it is straightforward to obtain analytically $E_{\text{ini}}$ and the width $\sigma_{\text{ini}}$ of the energy shell,
\begin{eqnarray}
\sigma_{\text{ini}} &=& \sqrt{\sum_{\alpha} |C_{\alpha}^{\text{ini}} |^2 (E_{\alpha} - E_{\text{ini}})^2}
=\sqrt{\sum_{n \neq \text{ini}} |\langle n |\widehat{H}_F | \text{ini}\rangle |^2 } \nonumber \\
&=& \frac{J}{2} \sqrt{M_1  + \lambda_F^2 M_2 } .
\label{deltaEm}
\end{eqnarray} 
Above $|n\rangle$ are the eigenstates of $\widehat{H}_\text{I}$ and the connectivity $M_1$ ($M_2$) corresponds to the number of states $|n\rangle$ directly coupled with $|\text{ini}\rangle$ via the NN (NNN) flip-flop term. The width $\sigma_{\text{ini}}$ depends only on the off-diagonal elements of $\widehat{H}_F $ written in the basis $|n\rangle$. When the initial state is a site-basis vector, these elements are not affected by $\Delta$ or the on-site energies. We may then have the same initial state evolving according to very different final Hamiltonians (with or without impurity, isotropic or anisotropic) and showing a very similar fidelity decay. 

In the case of the N\'eel state, the picture is yet more general, because $M_2=0$, so $\sigma_{\text{ini}}$ does not even depend on $\lambda_F$. For this state, the only distinction caused by different $\widehat{H}_\text{F}$'s is on the level of saturation~\cite{Torres2014PRA,Torres2014NJP}. If the parameters of the final Hamiltonian take $E_{\text{ini}}$ closer to the edge of the spectrum, where the states are more localized and $\text{IPR}_{\text{ini}}$ is smaller, the saturation point is higher [see values of $\text{IPR}_{\text{ini}}$ in Table~\ref{table:ipr} of Sec.~\ref{Sec:Fluctuations}].

The Gaussian decay of the fidelity until saturation for the N\'eel state for integrable, chaotic, isotropic, anisotropic, clean and disordered final Hamiltonians is not an artifact of the system size. We confirmed it for $L$ up to 24 \cite{Torres2014PRA,Torres2014NJP}.

The Gaussian behavior until saturation is observed also for various other site-basis vectors evolving under various choices of parameters $\Delta_F, \lambda_F$, and $d_F$. The same happens also for initial states from different initial Hamiltonians, such as XX and XXZ models~\cite{Torres2014NJP}. In Figs.~\ref{fig:LDOS} (e) and (f), we show $P^{\text{ini}}_{G}(E) $ and $F_{\text{G}}(t)$, respectively, for an initial state from the XX model with $E_{\text{ini}}$ close to the middle of the spectrum and evolving according to a chaotic Hamiltonian similar to the one considered for the N\'eel state in Figs.~\ref{fig:LDOS} (c) and (d).  The Gaussian behavior again persists all the way to saturation. Notice also that the time to reach the saturation point in Fig.~\ref{fig:LDOS} (f) is longer than for the N\'eel state, happening now at $t_R\sim 2 J^{-1}$.

There are cases, however, where despite the strong coupling regime, the fidelity decay transitions from Gaussian to  exponential before saturation. In some of these cases, this is associated with the poor filling of the energy shell, which is expected. But for others, the filling is not that bad, or is at least comparable to that of the N\'eel state under integrable Hamiltonians. It is not very clear yet what causes this transition in our systems and how to calculate the critical time $t_c$ where it happens. A way to estimate $t_c$ for two-body- and band-random matrices was suggested in \cite{Flambaum2001a,Flambaum2001b,Izrailev2006}, but by following those steps we obtained a broad range of values for $t_c$. This subject still requires further analysis.

%%%%%%%%%%%%%%%% OBSERVABLES %%%%%%%%%%%%%%%%%%%%%%%%%
\subsection{Few-Body Observables}
\label{Sec:Obs}

The evolution of few-body observables $O$ depends on more factors than the fidelity decay, but a simple general picture, valid at short times, can be constructed when the operator $\widehat{O}$  commutes with $\widehat{H}_I$. In this case, the dynamics is given by
\begin{equation}
O(t) = F(t) O(0) 
+\sum_{n\neq \text{ini}}  O_{n, n} \left| \langle n | e^{- i \widehat{H}_F t} | \text{ini} \rangle  \right|^2 .
\label{eq:TotObs}
\end{equation}
At short times, the expansion of Eq.~(\ref{eq:fidelity}), independently of the shape of the LDOS, leads to $F(t)\approx 1-\sigma_{\text{ini}}^2 t^2 $. As a result, the second order expansion of $O(t)$ simplifies to
\begin{equation}
O(t) \approx \left( 1-\sigma_{\text{ini}}^2 t^2 \right)  O(0) + t^2 \sum_{n \neq \text{ini}} | \langle n | \widehat{H}_F  | \text{ini} \rangle |^2 O_{n, n}.
\label{eq:ObsCommute}
\end{equation}
This quadratic behavior in $t$ is not necessarily obtained when $\widehat{O}$ does not commute with the initial Hamiltonian~\cite{Torres2014NJP}.

When the initial state is a site-basis vector, all the observables oriented along the $z$ direction satisfy Eq.~(\ref{eq:ObsCommute}). Many are experimentally accessible. They include the on-site magnetization, $\widehat{S}_j^z$;  the spin-spin correlation in the $z$ direction between sites $i$ and $j$, 
\begin{equation}
\widehat{C}^{z}_{i,j} (t)=  \widehat{S}^z_{i} \widehat{S}^z_{j} ,
\label{cxx} 
\end{equation}
and the structure factor in the $z$ direction,
\begin{equation}
\widehat{s}_f^{z}(\kappa ) = \frac{1}{L} \; \sum_{j,k=1}^L e^{-i \kappa (j-k)} \;
\widehat{S}_j^{z} \; \widehat{S}_k^{z} ,
\end{equation}
where $ \kappa=2\pi p/L$ stands for momentum and $p=0,1,2 \ldots, L$ is a positive integer. 

For site-basis vectors, Eq.~(\ref{eq:ObsCommute}) does not depend on $\Delta_F$ or $d_F$, and for the N\'eel state not even on $\lambda_F$. The short-time dynamics of the observables can then be equivalent despite being governed by very different $\widehat{H}_F$'s. For the N\'eel state, Eq.~(\ref{eq:ObsCommute}) for the three observables above gives the very simple expressions,
\begin{eqnarray}
&&S_{L/2}^{z,|\rm{NS}\rangle}(t)  \approx  S^z_{L/2}(0)  \left[1 -  J^2 t^2 \right],   \\
&&C^{z,|\rm{NS}\rangle}_{\frac{L}{2},\frac{L}{2}+1}(t)  \approx  -\frac{1}{4} \left[1 -  J^2 t^2 \right],  \label{eq:Cns} \\
&&s_f^{z,|\rm{NS}\rangle}(\pi,t) \approx \frac{L}{4} - \frac{J^2 t^2}{2} \left( \frac{2}{L} - 3 +L \right). \label{eq:sz} 
\end{eqnarray}
Notice that the structure factor, which is a nonlocal observable in position, shows a dependence on $L$ that is absent in the local observables. 

\begin{figure}[htb]
\centering
\includegraphics*[width=3.4in]{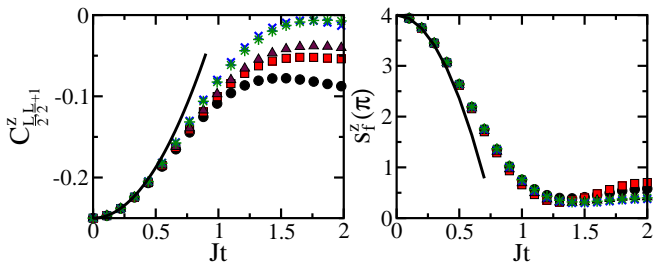}
\caption{(Color online) Evolution of the spin-spin correlation (left) and structure factor (right), both in the $z$ direction, for the N\'eel state.
The final Hamiltonians are clean, $\varepsilon=d_F=0$, and have the following parameters: $\Delta_F=1,\lambda_F=0$ (circle),  $\Delta_F=0.5,\lambda_F=0$ (square), $\Delta_F=1,\lambda_F=0.4$ (triangle), $\Delta_F=1,\lambda_F=1$ (cross), and $\Delta_F=0.5,\lambda_F=1$ (star). Solid curves are the analytical results from Eq.~(\ref{eq:Cns}) and (\ref{eq:sz}); $L=16$.}
\label{fig:czzns}
\end{figure}

The comparison between numerical results and the analytical expressions (\ref{eq:Cns}) and (\ref{eq:sz}) for five different clean final Hamiltonians are shown in Fig.~\ref{fig:czzns}. The agreement at short times is very good. The results confirm that the dynamics is independent of $\Delta_F$ or $\lambda_F$, all curves falling on top of each other. 

The saturation value of  $C^{z}_{\frac{L}{2},\frac{L}{2}+1}$ on the left panel of Fig.~\ref{fig:czzns}  is closest to zero when $E_{\text{ini}}$ is closest to the center of the spectrum. Among the final Hamiltonians considered, this happens for the strongly chaotic and anisotropic $\widehat{H}_F$ ($\Delta_F=0.5,\lambda_F=1$). Scaling analysis performed in~\cite{Torres2014NJP} suggests that in the thermodynamic limit, the saturation value indeed goes to zero for $\widehat{H}_F$ in the chaotic regime. In the integrable domain the results indicate a value different from zero and therefore possible memory retainment.

The short-time expressions obtained here for experimental observables  constitute an important first step towards the derivation of more general analytical expressions for the entire relaxation process. Long-time analytical expressions have been obtained for the evolution of the Shannon entropy~\cite{Santos2012PRL,Santos2012PRE} and may shed light on future studies about the observables.

%%%%%%%%%%%%%%%%%%%% FLUCTUATIONS %%%%%%%%%%%%%%%%%%%%%
\section{Time Fluctuations}
\label{Sec:Fluctuations}

In isolated quantum systems, the relaxation process is due to dephasing. The system will have relaxed to a new equilibrium, if after the transients have died, there are only fluctuations around a steady state. The fluctuations need to be small and decrease with system size, so that they disappear in the thermodynamic limit. 

Based on semiclassical arguments and full random matrices~\cite{Feingold1986,Deutsch1991,Prosen1994,SrednickiARXIV,Srednicki1996,Srednicki1999}, it was shown that the temporal fluctuations of few-body observables, $\sigma_O$, decrease exponentially with system size. 
In contrast, in the case of a non-interacting integrable Hamiltonian or mapped onto one, it was shown analytically~\cite{Venuti2013} and numerically~\cite{Cassidy2011,Gramsch2012,HeSantos2013} that the time fluctuations of one-body or quadratic observables scale as $1/\sqrt{L}$. 

This made us wonder whether the differences were indeed associated with the regime (chaotic or integrable) of the system, as the results suggested, or were caused by something else. We focused on realistic systems with two-body interactions, integrable and chaotic, and analyzed how $\sigma_O$ scales with $L$ and how it depends on the energy of the initial state. Our numerical results for spin-1/2 models showed that {\bf \em the decay of the temporal fluctuations of few-body observables with system size is exponential for both integrable and chaotic systems, provided interactions be present and the energy of the initial state is not too close to the edge of the spectrum~\cite{Zangara2013}}. 

These findings are in agreement with~\cite{Reimann2008,Short2011,Short2012,Reimann2012}, where analytical studies for the upper bounds of the variance, $\sigma_O^2$,  was shown to be given by
\begin{equation}
\sigma^2_O \leq \frac{(O_{\text{max}} - O_{\text{min}})^2 }{\text{IPR}_{\text{ini}} } ,
\label{bound}
\end{equation}
where $O_{\text{max(min)}}$ is the maximum (minimum) eigenvalue of the operator $\widehat{O}$. This result is valid for any initial state delocalized in the energy eigenbasis of any final Hamiltonian without too many degeneracies of eigenvalues and energy spacings. If the initial state is not too close to the edge of the spectrum, IPR$_{\text{ini}}$ should grow exponentially with $L$. Moreover, we verified that systems with interactions, even when integrable, are indeed not highly degenerate.  

In general, the distribution $P(s)$ of spacings $s$ between neighboring levels of integrable models is Poisson. This distribution indicates the presence of degenerate eigenvalues, but also the existence of many non-degenerate eigenvalues. For the interacting integrable XXZ model, this is the observed distribution. In this case, we confirmed numerically that the number of degenerate spacings between any two eigenvalues is also limited. In contrast, for the XX Hamiltonian (non-interacting integrable model), $P(s)$ is simply a peak at zero spacing, indicating an enormous amount of degenerate eigenvalues. This also leads to a large number of degenerate level spacings. 

As for the dependence on $E_{\text{ini}}$, the coefficient $\kappa$ of the exponential decay, $\sigma_O \propto \exp(-\kappa L)$, becomes smaller as the energy of the initial state moves away from the center of the spectrum and $\text{IPR}_{\text{ini}} $ decreases, but overall exponential fittings are better than power-law~\cite{Zangara2013}.

Here, we illustrate the exponential decay of the fluctuations with $L$ for the fidelity. Using Eqs.~(\ref{eq:Fevolved}) and (\ref{eq:Fsaturated}), the
variance of the temporal fluctuations of $F(t)$ is
\begin{eqnarray}
&&\sigma^2_F = \overline{ |F(t)  - \overline{ F(t) }|^2}  
\nonumber  \\
&&= 
\mathop{\sum_{\alpha \neq \beta} }_{\gamma \neq \delta} 
|C_{\alpha}^{\text{ini}}|^2  |C_{\beta}^{\text{ini}}|^2 |C_{\gamma}^{\text{ini}}|^2  |C_{\delta}^{\text{ini}}|^2
\overline{e^{i (E_{\alpha} - E_{\beta}+E_{\gamma} - E_{\delta}) t}} .
\nonumber
\end{eqnarray}
The exponential averages out, unless $E_{\alpha} - E_{\beta} = E_{\delta} - E_{\gamma} $. Under the condition of non-degenerate energy spacings, this implies  $E_{\alpha} = E_{\delta}$ and $E_{\beta} = E_{\gamma} $, which leads to
\begin{eqnarray}
\sigma^2_F &=& 
\sum_{\alpha} 
|C_{\alpha}^{\text{ini}}|^4 \sum_{\beta}  |C_{\beta}^{\text{ini}}|^4 - \sum_{\alpha} 
|C_{\alpha}^{\text{ini}}|^8
\nonumber \\
&=& \frac{1}{\text{IPR}_{\text{ini}}^2 } - \sum_{\alpha} 
|C_{\alpha}^{\text{ini}}|^8 .
\label{Eq:sigmaF}  
\end{eqnarray}
When the initial state is substantially delocalized and ${\cal D}$ is large, 
\[
\left(\sum_{\alpha} 
|C_{\alpha}^{\text{ini}}|^4\right)^2 \gg  \sum_{\alpha} 
|C_{\alpha}^{\text{ini}}|^8 \Rightarrow \sigma_F \sim \overline{F},
\] 
the standard deviation $\sigma_F$ coincides with the infinite time average of the fidelity.

\begin{figure}[htb]
\centering
\includegraphics*[width=2.2in]{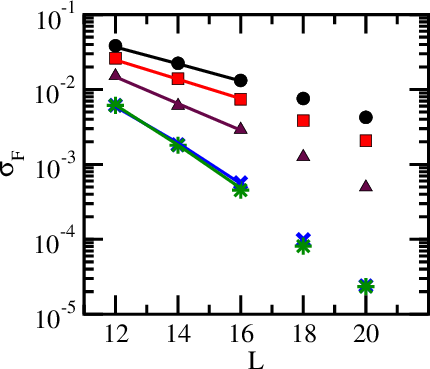}
\caption{(Color online) Logarithmic plot of the standard deviation
of the time fluctuations of the fidelity  vs $L$ for the N\'eel state.
The final Hamiltonians are clean, $\varepsilon=d_F=0$, and have the following parameters: $\Delta_F=1,\lambda_F=0$ (circle),  $\Delta_F=0.5,\lambda_F=0$ (square), $\Delta_F=1,\lambda_F=0.4$ (triangle), $\Delta_F=1,\lambda_F=1$ (cross), and $\Delta_F=0.5,\lambda_F=1$ (star). Solid curves are the analytical results from Eq.~(\ref{Eq:sigmaF}).}
\label{fig:fluctuations}
\end{figure}

In Fig.~\ref{fig:fluctuations}, we show the standard deviation of the time fluctuations of the fidelity for the N\'eel state. Symbols are numerical results and solid lines correspond to the results from Eq.~(\ref{Eq:sigmaF}). Notice that the latter requires exact diagonalization, which we can perform for $L$ up to 16. We see from the figure that for the interacting systems, in the integrable ($\lambda_F=0$) or chaotic ($\lambda_F>0$) domains, $\sigma_F$ decays exponentially with $L$. The value of the exponent of this decay increases with $\text{IPR}_{\text{ini}}$. For the N\'eel state, $\text{IPR}_{\text{ini}}$ increases with $\lambda_F$ and decreases with $\Delta_F$  (cf. Fig.~\ref{fig:fluctuations} and Table~\ref{table:ipr}).

\begin{table}[h]
\caption{$E_\text{ini}$ and $\text{IPR}_\text{ini}$ for the N\'eel state; $\varepsilon=0$, $d=0$, $L = 16$.}
\begin{center}
\resizebox{5.5cm}{!}
{
\begin{tabular}{ccl}
\hline \hline
\hspace{0.3 cm}$\widehat{H}_F$&\hspace{0.4 cm}$\text{E}_\text{ini}$&\hspace{0.4 cm}$\text{IPR}_\text{ini}$\\ 

\hspace{0.3 cm}$\Delta_F=1.0,\lambda_F=0.0$&\hspace{0.4 cm}-3.750&\hspace{0.4 cm}$72.15$\\ 
 \hspace{0.3 cm}$\Delta_F=0.5,\lambda_F=0.0$&\hspace{0.4 cm}-1.875&\hspace{0.4 cm}$129.83$ \\ 
 \hspace{0.3 cm}$\Delta_F=1.0,\lambda_F=0.4$&\hspace{0.4 cm}-2.350&\hspace{0.4 cm}$336.78$\\ 
 \hspace{0.3 cm}$\Delta_F=1.0,\lambda_F=1.0$&\hspace{0.4 cm}-0.250&\hspace{0.4 cm}$1805.25$\\ 
 \hspace{0.3 cm}$\Delta_F=0.5,\lambda_F=1.0$ &\hspace{0.4 cm}-0.125&\hspace{0.4 cm}$2071.92$\\ 
\hline
\hline
\end{tabular}
}
\end{center}
\label{table:ipr}
\end{table}

For the strongly chaotic final Hamiltonians, $\Delta_F=1.0,\lambda_F=1.0$ and $\Delta_F=0.5,\lambda_F=1.0$, fittings to the curves of Fig.~\ref{fig:fluctuations} indicate that the coefficient $C$ in $\sigma_F \propto \exp(-C L)$ is $C\sim 0.7$. This value coincides with the decay expected for an initial state that corresponds to an eigenstate of a GOE full random matrix, where $\text{IPR}_\text{ini} \sim {\cal D}/3$. For the subspace considered here,  ${\cal D} = L!/(L/2)!^2$. Using Stirling's approximation, the GOE scenario leads to 
\[
\ln \sigma_F \sim \ln(3/{\cal D})=\ln3 -L \ln2 .
\]
The agreement between the values of $C\sim \ln2 \sim 0.7$ suggests that the N\'eel state projected on strongly chaotic two-body-interaction Hamiltonians behaves as a chaotic state from GOE matrices.

%%%%%%%%%%%%%%%%%%%% THERMALIZATION %%%%%%%%%%%%%%%%%%%%%
\section{Thermalization}
\label{Sec:Thermalization}

After relaxation, a natural question is whether the observable reaches thermal equilibrium or not.
In isolated many-body quantum systems, the system can play the role of an environment to few-body observables, allowing them to thermalize. 

The dynamics of the expectation value of an observable is given by
\begin{eqnarray}
&&O(t)
=\langle\Psi(t)|\widehat{O}|\Psi(t)\rangle
\label{eq:Oboth} \\
&&=\sum_{\alpha} |C_{\alpha}^{\text{ini}}|^2 O_{\alpha \alpha} + \sum_{\alpha \neq \beta} C_{\alpha}^{\text{ini*}} C_{\beta}^{\text{ini}} e^{i(E_\alpha-E_\beta)t} O_{\alpha \beta}\;, \nonumber
\end{eqnarray}
where $O_{\alpha \beta} = \langle\psi_\alpha|\widehat{O}|\psi_\beta \rangle$. In the absence of too many degeneracies, the off-diagonal elements of $O(t)$ oscillate very fast and cancel out on average, so the infinite time average is 
\begin{equation}
   \overline{O} =\sum_\alpha |C_{\alpha}^{\text{ini}}|^2 O_{\alpha\alpha} .
\label{eq:Ode}
\end{equation}
We note that in generic systems, the eigenstate expectation values $O_{\alpha \alpha} = \langle\psi_\alpha|\widehat{O}|\psi_\alpha \rangle$ are usually larger than $O_{\alpha \beta}$, which also guarantees negligible contributions of the second term in Eq.~(\ref{eq:Oboth}) \cite{SrednickiARXIV,Santos2010PREb}.

The observable thermalizes when its infinite time average coincides with its thermal (microcanonical) average,
\begin{equation}
O_\text{ME}  \equiv  \frac{1}{{\cal{N}}_{E_{\text{ini}},\delta  E}}\hspace{-0.5cm}\sum_{\substack{\alpha \\ |E_{\text{ini}}-E_\alpha|<\delta E}}\hspace{-0.5cm} \langle \psi_{\alpha} |\widehat{O} | \psi_{\alpha } \rangle ,
\label{eq:micro}
\end{equation}
where ${\cal{N}}_{E_{\text{ini}},\delta E}$ stands for the number of energy eigenstates in the window $\delta E$. The equality between the two averages can only hold in the thermodynamic limit. When dealing with finite systems, which is the case experimentally, they can only be similar. The essential question is then when the two averages are close and whether they further approach each other as $L$ increases.

The proximity of the two averages have been associated with two scenarios:

(1) The eigenstate expectation value of the observables, $O_{\alpha \alpha}$, is a smooth function of energy. This means that the result from a single eigenstate inside the microcanonical window agrees with the microcanonical average, which is basically a statement of the validity of statistical mechanics. This approach became known as eigenstate thermalization hypothesis (ETH) \cite{Deutsch1991,Srednicki1994,Rigol2008}.

(2) The components $|C_{\alpha}^{\text{ini}}|^2$ behave as Gaussian random variables. This happens when the Gaussian LDOS fills the energy shell ergodically and the initial state has a very large temperature. In this framework, thermal features emerge even when the final Hamiltonian is integrable, as discussed in  \cite{Torres2013,He2013}. 

Our main focus here is on the validity of case (1). It is trivially satisfied in the scenario of full random matrices, since all eigenstates are just random vectors and so all $O_{\alpha \alpha}$'s are equivalent. But our interest is on realistic systems with two-body interactions, where chaotic eigenstates can exist only away from the edges of the spectrum. In this case,  {\bf \em thermalization can happen if the energy of the initial state and the width of its LDOS fall inside the region of chaotic eigenstates of $\widehat{H}_F$}.

\subsection{Local and Global Quenches: Comparable Effects}

To further discuss the issue of thermalization, we consider two quenches that take the system into the chaotic domain.
We deal with the spin Hamiltonian~(\ref{ham}) in the subspace that has $L/3$ up-spins, implying dimension ${\cal D}=L!/[(2L/3)!(L/3)!]$ and ${\cal{S}}^z=-L/6$. Our system starts in an eigenstate of the initial Hamiltonian corresponding to the integrable XXZ model with a small defect on site 1 of amplitude $\varepsilon = 0.1 $:
\[
\widehat{H}_I = \varepsilon J \widehat{S}_{1}^z + \widehat{H}_{\text{NN}}.
\]
Two perturbations are then carried out:
\begin{itemize}
\item A local quench in space, where the perturbation is localized on a single site:  $d_\text{I}= 0 \rightarrow d_\text{F} \neq 0$. The final Hamiltonian becomes the chaotic impurity model with NN couplings only,
\[
\widehat{H}_F^{\text{local}} = \widehat{H}_I + d_F J \widehat{S}_{\lfloor L/2 \rfloor }^z .
\]
\end{itemize}
\begin{itemize}
\item A global quench in space, where the perturbation affects simultaneously all sites in the chain: $\lambda_\text{I}=0\rightarrow \lambda_\text{F} \neq 0$. The final Hamiltonian becomes the chaotic Hamiltonian with NNN couplings,
\end{itemize}
\[
\widehat{H}_F^{\text{global}} = \widehat{H}_I +  \lambda_F \widehat{H}_{\text{NNN}} .
\]

We have shown that, for the same initial states, both quenches may lead to very similar results, provided the perturbation is not too strong~\cite{Torres2014PRE}. This is because in this limit of intermediate perturbation, $\widehat{H}_F^{\text{local}}$ and $\widehat{H}_F^{\text{global}}$ written in the eigenstates of $\widehat{H}_I$ have very similar structures, resulting in equivalent signatures of chaos associated with eigenvalues and eigenstates.

\begin{figure}[htb]
\centering
\includegraphics*[width=2.in]{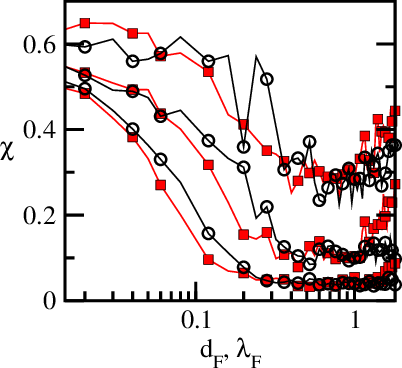}
\caption{(Color online) Indicator  $\chi$ of the integrable-chaos crossover vs the perturbation strength for the impurity (filled squares) and the NNN (empty circles) models; $\Delta_{F}=0.48$. From top to bottom: $L=12,15,18$.}
\label{fig:Ps}
\end{figure}
To quantify the crossover from integrability to chaos, we show in Fig.~\ref{fig:Ps}  the level spacing indicator $\chi$ defined as \cite{Santos2010PRE,Torres2014PRE}
\begin{equation}
\chi \equiv \frac{\sum_i[{\cal{P}}(s_i)-{\cal{P}}_{WD}(s_i)]}{\sum_i {\cal{P}}_{WD}(s_i)}.
\label{eta}
\end{equation}
Above, the sums run over the whole spectrum, $s$ is the spacing between neighboring unfolded energies, ${\cal{P}}(s)$ is the level spacing distribution, and ${\cal{P}}_{ WD}(s) = (\pi s/2)\exp(-\pi s^2/4)$ is the Wigner-Dyson distribution obtained for chaotic systems with real and symmetric Hamiltonians. For interacting integrable systems, the level spacing distribution is Poisson, ${\cal{P}}_{ P}(s) = \exp(-s)$. Close to the integrable domain, $\chi$ is therefore large and it approaches zero in the chaotic regime.

In Fig.~\ref{fig:Ps}, as  $d_F$ and $\lambda_F$ increase, both models become chaotic and show similar values of $\chi$ for the same system sizes. If the perturbation is further increased well above 1, the systems eventually reach another integrable point. Notice also that as $L$ increases, not only does $\chi$ decrease, but also the value of the perturbation leading to small $\chi$. The onset of chaos in the thermodynamic limit may therefore require an infinitesimally small integrability breaking term~\cite{Santos2010PRE,Torres2014PRE}.

\subsubsection{Chaotic Eigenstates and ETH}

\begin{figure}[h]
\centering
\includegraphics*[width=2.7in]{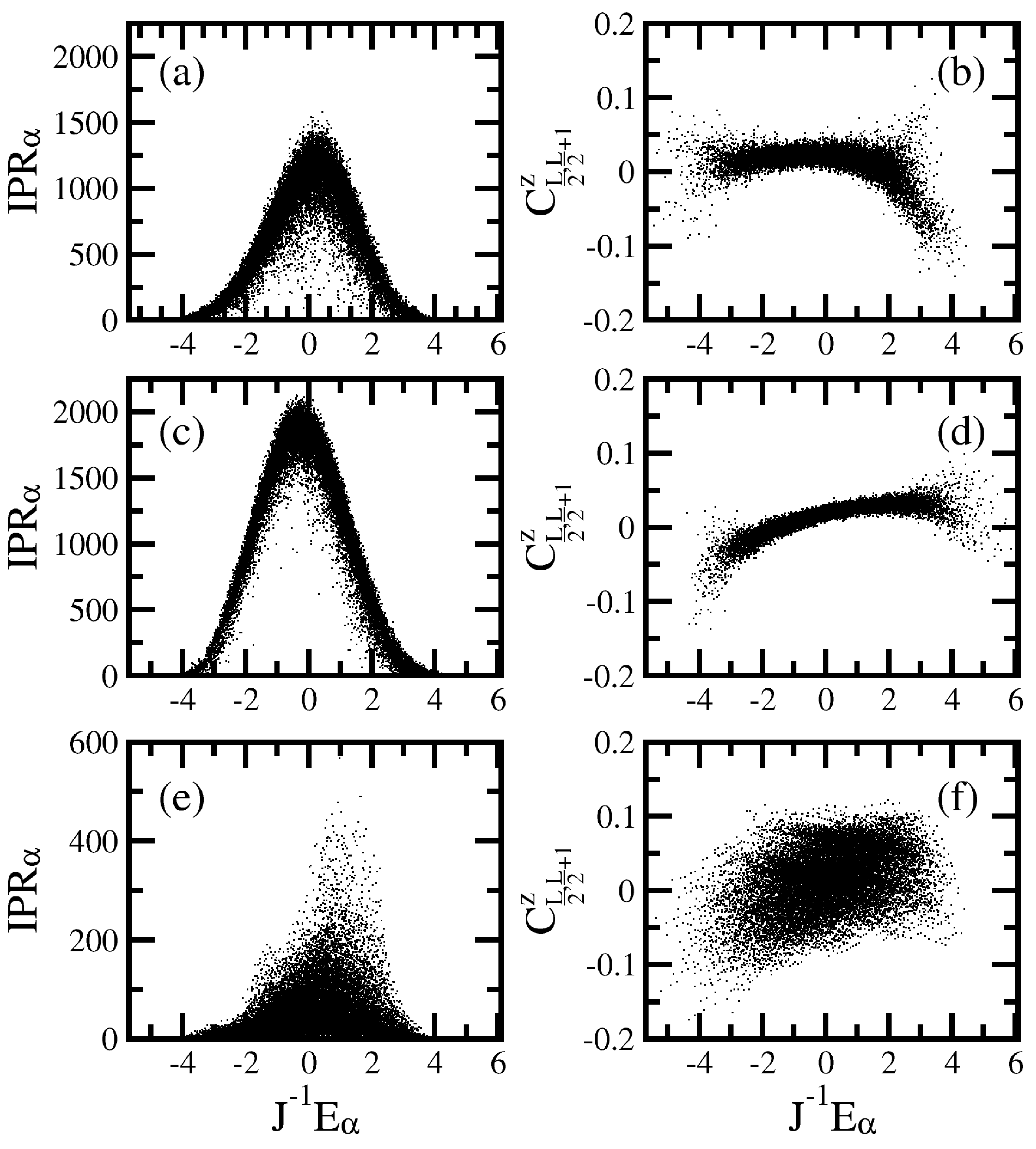}
\caption{(Color online) Inverse participation ratio (left) and expectation values of the spin-spin correlation (right) vs $E_{\alpha}$ for all the eigenstates of the Hamiltonians with $\Delta_F=0.48$ and $d_F=0.9$ (a,b); $\lambda_F=0.44$ (c,d); and $d_F=\lambda_F=0$ (e,f). IPR is computed in the basis corresponding to the eigenstates of the XXZ model with $\Delta_I=0.48$ (a,c) and the eigenstates of the XX model (e). For all cases: $\varepsilon=0.1$ and $L=18$. }
\label{fig:ETH}
\end{figure}

The eigenstates $|\psi_{\alpha}\rangle$ of a chaotic $\widehat{H}_F$ written in the eigenstates of $\widehat{H}_I$ and away from the edges of the spectrum are very delocalized~\cite{Santos2010PRE,RigolSantos2010,Santos2010PREb,Torres2014PRE}. They are close to random vectors, although they never reach the level of spreading of eigenstates from full random matrices. Their values of $\text{IPR}_{\alpha}$ are large and become smooth functions of energy, as seen in Fig.~\ref{fig:ETH} (a) and (c). This is to be contrasted with the integrable XXZ model, where large fluctuations prevail [Fig.~\ref{fig:ETH} (e)]. 

The similar structures of the eigenstates in the chaotic domain lead to small fluctuations of  $O_{\alpha \alpha}$. This is illustrated in Fig.~\ref{fig:ETH} (b) and (d) for the spin-spin correlation. For the parameters chosen in the figure, the values of $\text{IPR}_{\alpha}$ for the eigenstates of $\widehat{H}_F^{\text{local}}$ and $\widehat{H}_F^{\text{global}}$ in the XXZ basis are equivalent~\cite{Torres2014PRE}. In this case the sizes of the fluctuations for both models are also comparable. In contrast, for the integrable system, the fluctuations are  much larger  [Fig.~\ref{fig:ETH} (f)].
By comparing the eigenstate expectation values of few-observables for three system sizes (the ones available to exact diagonalization), we also verified that the fluctuations  decrease with system size for the two chaotic systems away from the edges of the spectrum~\cite{Santos2010PREb,Torres2014PRE}, but not for the integrable model or the chaotic ones close to the borders of the spectrum. These results indicate the viability of thermalization in the region of chaotic eigenstates.

\subsubsection{Infinite Time Average and Thermal Average}

Analyses of the sizes of the fluctuations of $\text{IPR}_{\alpha}$ and $O_{\alpha \alpha}$ for different $L$'s serve as indications for when thermalization may happen. The conclusions can be reinforced by comparing $\overline{O}$ and $O_{\text{ME}}$. To calculate the infinite time average, an initial state needs to be selected.  This is done by searching the $E_{\text{ini}}$ that is closest to the energy
\begin{equation}
E_T=\frac{\sum_\alpha E_\alpha e^{-E_\alpha/k_B T}}{\sum_\alpha e^{-E_\alpha/k_B T}},
\end{equation}
for a specific temperature $T$ ($k_B$ is Boltzmann constant). This procedure associates with the initial state an intensive quantity, which allows for a fair comparison between different systems  of different sizes.

To illustrate how the distance between the two averages depend on perturbation, energy of the initial state, and system size, we choose the spin-spin correlation in the $z$ direction and compute
\begin{equation}
\Lambda C^{z}_{\frac{L}{2},\frac{L}{2}+1}=\left| \overline{C^{z}_{\frac{L}{2},\frac{L}{2}+1} }- \left( C^{z}_{\frac{L}{2},\frac{L}{2}+1}\right)_{\text{ME}} \right| .
\label{eq:reldif}
\end{equation}

\begin{figure}[htb]
\centering
\includegraphics*[width=3.2in]{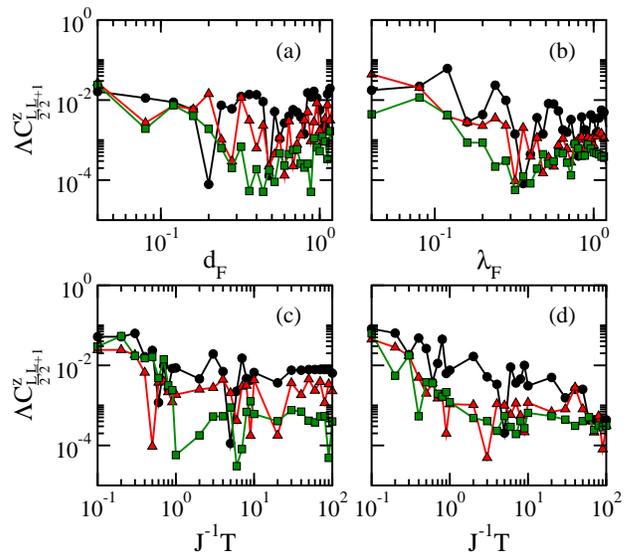}
\caption{(Color online) $\Lambda C^{z}_{\frac{L}{2},\frac{L}{2}+1}$ vs $d_F$ (a) and $\lambda_F$ (b) and  vs temperature (c,d). Top panels: Initial state with $k_B T=7 J$. Bottom panels: $d_F=0.9$ (c) and $\lambda_F=0.44$ (d). All panels: $\epsilon=0.1$, $\Delta_{I,F}=0.48$; $L=12$ (circles); $L=15$ (triangles);  $L=18$ (squares).}
\label{fig:thermalization}
\end{figure}

The top of Fig.~\ref{fig:thermalization} shows $\Lambda C^{z}_{\frac{L}{2},\frac{L}{2}+1}$ for different values of the perturbations for the quenches to the impurity (a) and NNN (b) Hamiltonians. $\Lambda C^{z}_{\frac{L}{2},\frac{L}{2}+1}$ is of similar magnitude for both models and clearly diminishes as the perturbation increases and the system becomes more chaotic. It reaches a minimum before increasing again due to the approach to a new integrable point (cf. Fig.~\ref{fig:Ps}). The difference between the two averages also decreases with $L$, which is a strong sign of the viability of thermalization.

The bottom of Fig.~\ref{fig:thermalization} shows that the averages improve as the temperature increases and  the energy of $|\text{ini}\rangle$ approaches the middle of the spectrum. The improvement with system size is also evident. These results corroborate the dependence on temperature in the studies of thermalization~\cite{He2013,Torres2013}. 

%%%%%%%%%%%%%%%%%%%% CONCLUSION %%%%%%%%%%%%%%%%%%%%%
\section{Conclusion}
\label{Sec:Summary}

This work overviews our recent studies of the behavior of isolated interacting quantum systems  from the moment they are taken out of equilibrium instantaneously until they reach a new equilibrium. We have employed one-dimensional spin-1/2 models, which are prototypes of many-body quantum systems. Only two-body interactions have been included. We summarize our results as follows.

(i) {\em Dynamics:} For realistic systems with two-body interactions, the probability of finding the initial state in time, the so-called fidelity, can show a Gaussian behavior until saturation~\cite{Torres2014PRA,Torres2014NJP,Torres2014PRAb}. This happens in the strong coupling regime, when the LDOS (weighted energy distribution of the initial state) is unimodal and has a Gaussian shape. In this case, the minimum time for the fidelity to reach the saturation point is $t_R=\sqrt{\ln(\text{IPR}_{\text{ini}}) }/\sigma_{\text{ini}}$. 

Faster than Gaussian decays require either the simultaneous interactions of more than two particles, the extreme being the case of full random matrices, or a bimodal (multimodal) LDOS. A GOE random matrix sets the lower bound for $t_R$ for initial states with single-peaked LDOS: $ [{\cal J}_1( 2 \sigma_{\text{ini}} t_R)]^2/(\sigma_{\text{ini}}^2 t_R^2) = 3/{\cal D}$.  The short-time decay of the fidelity for a bimodal LDOS can reach the ultimate lower bound, $F(t) \geq \cos^2(\sigma_{\text{ini}} t)$, established by the energy-time uncertainty relation~\cite{Torres2014PRAb}.

The analysis of the evolution of few-body observables is still under progress. We have so far developed a simple picture for the observables that commute with the initial Hamiltonian. In this case, the short-time dynamics is $\propto t^2$. For initial states corresponding to site-basis vectors it is straightforward to derive analytical expressions for experimental observables, such as magnetization and  spin-spin correlation~\cite{Torres2014PRA,Torres2014NJP}. 

(ii) {\em Fluctuations after Relaxation:} Overall the time fluctuations after relaxation decay exponentially with system size, even when the system is integrable, provided it does not have a large number of degeneracies~\cite{Zangara2013}. The size of the fluctuations depends on the observables and on the level of delocalization of the initial state in the energy eigenbasis. For the fidelity, the standard deviation is $\sigma_F = \sqrt{ \text{IPR}_{\text{ini}}^{-2}  - \sum_{\alpha} 
|C_{\alpha}^{\text{ini}}|^8}$.

(iii) {\em Thermalization:} In an isolated many-body quantum system, a few-body observable can thermalize if the energy of the initial state and the width of its LDOS fall inside the region of chaotic eigenstates of the final Hamiltonian~\cite{Santos2010PRE,Torres2013,Torres2014PRE}. It is in this chaotic region of the spectrum that notions of statistical mechanics, and equivalently ETH or typicality, can hold. In this region, the eigenstate expectation values of few-body observables do not fluctuate much for eigenstates close in energy, so the result from one eigenstate agrees with the result from the average. These fluctuations decrease as the system size increases, leading to the coincidence of the infinite time average and the microcanonical average in the thermodynamic limit. 

As illustrations, we considered here local and global quenches in space that take the system into the chaotic domain. The difference between the infinite time average and the microcanonical average decreases as the perturbation increases and the eigenstates of $\widehat{H}_F$ become more chaotic, as the initial state moves toward the middle of the spectrum where chaotic states reside, and as the system size increases~\cite{Torres2014PRE}. 

In the thermodynamic limit, it has been shown that when the energy of the initial state is not very close to the middle of the spectrum, integrable systems cannot thermalize~\cite{Rigol2014}. It remains to elucidate where the energy threshold is located. How close to the middle of the spectrum does the initial state need to be for thermal behavior to be possible in integrable systems? How close to the borders of the spectrum of a gapless chaotic system can the initial state be for thermalization to occur?

Another open question refers to the level of chaoticity of the eigenstates in real systems. They are never as much spread as the eigenstates of full random matrices. It remains to understand the effects that this can have on different observables.

%%%%%%%%%%%%%%%%%%%% ACKNOWLEDGMENTS %%%%%%%%%%%%%%%%%%%%%
\begin{acknowledgments}
This work was supported by the  NSF grant No.~DMR-1147430. E.J.T.H. acknowledges partial support from CONACyT, Mexico.
\end{acknowledgments}

\end{document}